# Hetero-interface of electrolyte/2D materials

**Xin Hu, Shou-Xin Zhao, Yang Li\*, Zhao-Yuan Sun, Liang Zhen, Cheng-Yan Xu \***

X. Hu, S.-X. Zhao, L. Yang, Z.-Y. Sun, L. Zhen, C.-Y. Xu
School of Materials Science and Engineering
Harbin Institute of Technology
Harbin 150001, China
E-mail: liyang2018@hit.edu.cn

Y. Li, L. Zhen, C.-Y. Xu
MOE Key Laboratory of Micro-Systems and Micro-Structures Manufacturing
Harbin Institute of Technology
Harbin 150080, China
E-mail: cy_xu@hit.edu.cn

Z.-Y. Sun
Center for Analysis and Measurement
Harbin Institute of Technology
Harbin 150001, China

L. Zhen, C.-Y. Xu
Sauvage Laboratory for Smart Materials
School of Materials Science and Engineering
Harbin Institute of Technology (Shenzhen)
Shenzhen 518055, China

## Abstract

Electrochemical gating has been demonstrated as a powerful tool to tune the physical properties of two-dimensional (2D) materials, leading to lots of fascinating quantum phenomena. However, the reported liquid-nature electrolytes (e.g., ionic liquid and ion-gel) cover the top surface of 2D materials, introduce the strain at the hetero-interface, and present sensitivity to humidity, which strongly limits the further exploration of the hetero-interface between electrolyte and 2D materials, and their wide applications for electronics and optoelectronics. Herein, by introducing a lithium-ion solid-state electrolyte, the character of the electric double layer (EDL) at hetero-interface and its effect on the optical property of transition metal chalcogenides (TMDs) have been revealed by Kelvin probe force microscopy (KPFM) and (time-resolved) photoluminescence measurements. The work function of TMDs can be strongly tailored by electrochemical gating, up to 0.7eV for WSe$_2$ and 0.3 eV for MoS$_2$, respectively. Besides, from the gate-dependent surface potential of TMDs with different thicknesses, the potential drop across the EDL has been quantitatively revealed. Furthermore, from the gate-dependent PL emission at room temperature, monolayer $WS_2$ exhibits only neutral exciton emission in the whole range of gate voltage applied, which also exhibits exciton-exciton annihilation with ($1.99 \pm 0.27 \times 10^{-3}$ $cm^2$ / s). Our results demonstrate that lithium-ion substrate is a promising alternative to explore the physics of 2D materials and the hetero-interface of electrolyte/2D materials by easily integrating both scanning probe and optical techniques.

1. **Introduction**

2D semiconductors have shown a variety of novel properties, such as high carriers mobility, spin valley polarization, and quantum effect on account of its distinctive nanoscale dimension. Transition metal dichalcogenides (TMDs) are the most developed and applicable 2D materials because its tunable optical band gap, weak van der Waals force, and dangling bond. Multiple electronic and optoelectronic devices, such as superconductivity devices, flexibility and wearable devices, biosensors, and synapse device, have been designed based on TMDs. These devices are usually gated by traditional oxide, like $SiO_2$, $Al_2O_3$ and $HfO_2$[1]. However, the oxide gate always requires high on-state voltage, thus leads to high energy consumption and the devices are easily to be breakdown.

In recently years, ion gel and ion liquid have been used to replace the $SiO_2$. These ion electrolyte shows strong gating ability through its functional structure of electric double layer (EDL). By extra electric field, ions are drove to drift along the direction of electric field line and ordering at interface to form such EDL. This structure has been used in a amount of designs of high performance device and has shown great potential both in various electrochemical devices and electric devices, such as lithium batteries, biosensing probe, neural synaptic, superconductivity FET, and metal-insulator transformation[1-6]. However, since the ion electrolyte covers the channel surface, there still lacks the knowledge of EDL interface and inner dynamic property against extra stimuli[3, 4, 7]. There is little visual and valid experimental description of local fluctuations of EDL interface, such as immigration and distribution of ions, charge transfer through the interface, the interaction between charged carriers. Solid lithium-ion glass substrate (SLGC) is another type ion electrolyte and is suitable to be treated as bottom gate[8-11]. Therefore, we choose solid electrolyte in order to expose surface of EDL, and such a device configuration provide an efficient platform to combine any other surface measurement technique.

In this work, we take full advantage of the solid electrolyte and combine it with other scanning experimental techniques to explore the nature of EDL and the charge transfer and photonic dynamics of TMDs transistors. It is worth noting that bipolar TMD materials, such as $WSe_2$, should exhibit more meaningful electronic dynamic properties because of electron and hole injection. Hence, we took $WSe_2$ nanoflake as research object. And we design monolayer $WS_2$ which has high quantum yield as experimental object and identify the photonic dynamics of EDL interface. The PL spectra would be used to give amount of information about charged carriers creation, mobility and recombination.

## 2. Result and discussion

We apply KPFM to gated $WSe_2$ nanoflake and quantitatively depict the surface potential distribution and EDL formation. The gate voltage swept from -1V to +1V, with sweeping rate is 0.1 V• $s^{-1}$. Figure 1b presents atomic force microscopy (AFM) topography of mechanically exfoliated $WSe_2$. The corresponding height profile shows some wrinkles, which is unavoidable on account of PDMS dry transfer procedure. In that case, the monolayer $WSe_2$ (light yellow dotted line in Fig. 1b) is certified by PL spectroscopy (Supply Information). Given that the thickness of monolayer $WSe_2$ is 0.8nm, the relative thickness between light yellow dotted line and cyan dotted line (blue dotted line) is 1.12nm (6.56nm), which correspond to 2-layer (10-layer). In Figure 1c~f, we performed gate-dependent KPFM analysis. The Kelvin Probe Force microscopy (KPFM) technique is routinely used for the precise measurement of the surface potential with a nanoscale lateral resolution was intended to detect the formation and distribution of EDL[12, 13]. It is evident that surface potential mapping of $WSe_2$ is homogeneous across the entire area within the experimental voltage range. Also, it might suggest that ions distribution of EDL is uniform at nanoscale and support the formation of EDL in some extent. Besides, at room temperature, this type device is stable and the morphology of WSe$_2$ keep all the same after several cycle. The response of WSe$_2$ shows no hysteresis with reverse voltage step and the performance is as well as the first cycle. That indicate that solid lithium glass ceramic is more reliable than other ion gel electrolyte. The colour of the measured surface potential mapping differs between Au electrode, monolayer WSe$_2$ nanoflake and several layer WSe$_2$ nanoflake and it becomes more prominent with positive gate voltage. It suggests that the surface potential of nanoflake increase with positive voltage and we attribute it to strong electron doping origin form EDL at interface. Besides, the surface potential varies from monolayer WSe$_2$ to several layer WSe$_2$ nanoflake, which is much smaller in thick area. The thickness dependency reflects that the surface potential can be screened by thicker WSe$_2$ nanoflake which is in agreement with Debye theory, and we deduce that the effective range of EDL is limited and about several nanometers.

We denote potential difference between Au electrode and WSe$_2$ nanoflake as $\Delta V_{CPD}$ :

$$e\Delta V_{CPD} = \Phi_{sample} - \Phi_{electrode},$$

where $\Phi_{sample}$ and $\Phi_{electrode}$ are work function of Au and WSe$_2$ nanoflake respectively.

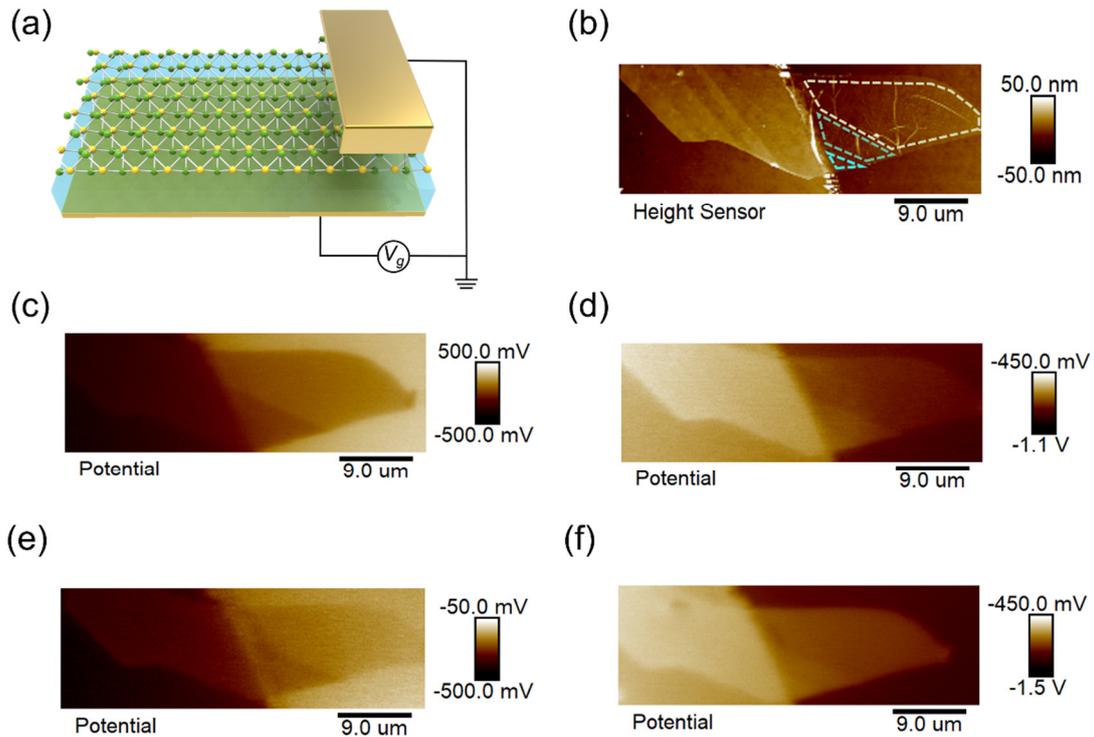

**Figure 1.** (a)Schematic diagram of WSe$_2$ device for surface potential analysis. Gate voltage is applied to top Au electrode and solid electrolyte substrate is grounded. (b) AFM topography of the device, the selected area with yellow,

cyran and blue line refers to thickness of monolayer, 2 layer and ~10 layers. (c) KPFM images with gate voltage of 1.0V (d) 0.3V (e) -0.3V (f) -1.0V

Given that work function of metal is a constant value and experimental tip condition keeps all the same, the variation of $\Delta V_{CPD}$ equals to work function variation of WSe$_2$ nanoflake. As shown in Fig 2a, the statistical thickness-dependent surface potential difference ($\Delta V_{CPD}$) is depicted as a function of gate voltage ($V_g$) by calibrating the potential profile in KPFM measurement. The surface potential of WSe$_2$ nanoflake shows a positive linear correlation with gate voltage, and the overall ascend is about 0.7V (monolayer), 0.54V (bilayer), 0.43V (~10 layers), 1.09V (Substrate), when the gate voltage varies from -1V to +1 V. Considering the bandgap of monolayer WSe$_2$ (1.63 eV), the EDL presents strong electron doping effect comparing with any traditional SiO$_2$. It is clear that solid electrolyte is proposed to be an ideal substrate to quantitively record surface electric property in EDL FET with significantly improved performance. Furthermore, Fig 2b present that the work function variation of WSe$_2$ decrease exponentially with the increase of layer number, and keeps constant (0.6V) when the layer number exceed 5-layers. It is noticeable that the sharp decrease of surface potential is confined within the thickness of ~0.8nm. We attribute such phenomenon to effective Debye length of EDL. It is depicted in Fig 2c that the lithium ions will drift along direction of electric field, and finally assembled at interface surface. Such assembled ions consist a parallel capacitor with several nanometers gap, intriguing significantly strong potential drop at surface. As the thickness of nanoflake increased to specific value, the surface potential can not be significantly influenced by interface high electric field and exhibit subtle surface potential variation with gate voltage. The above phenomenon is in well consistent with the effective thickness of EDL(~0.8nm).

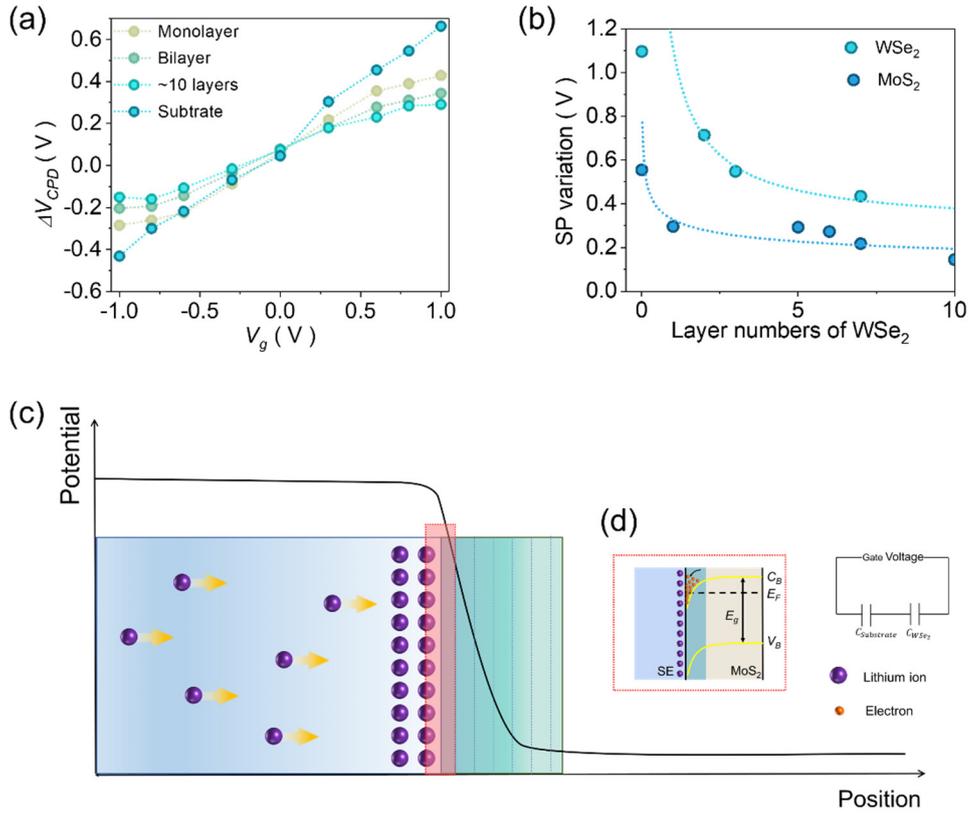

**Figure 2.** (a)The contact potential difference between the Au electrode and WSe₂ nanoflake as a function of gate voltage. (b) The contact potential difference as a function of thickness with gate voltage of 1.0V. (c) The schematic diagram of formation of EDL, the spatial distribution of mobile ions and the uniform experimental result of surface potential. (d) Band diagram across the device thickness.

Monolayer $WS_2$ is a typical direct-gap semiconductor showing strong photoluminescence quantum yield through visible spectral range. The device configuration is similar to Fig 1a. In Fig 3a, we performed temperature-dependent PL spectral of monolayer $WS_2$ with temperature varying from 273K to 77K. In order to reduce the possible thermal damage to monolayer $WS_2$, we kept the laser excitation density a constant as low as $1.21 \mu J \cdot cm^{-2}$ and make sure that heating effect is negligible under this excitation condition. There exists only one measurable PL peak positioned at 1.98 eV at room temperature as shown in Fig 3a and we recognized it as exciton (X) peak with reference to published peak position in recent researches. However, the exciton peak position is slightly smaller than others, and we suggest that the strong interaction of EDL goes for subtle blue shift of exciton peak. As depicted in Fig 3a, the exciton peak position experiences a blue shift with temperature decreasing to 77K. The change of exciton binding energy accounts for the phenomenon of peak shifting, which has been quantitively testified by Varshin theory. As discussed before, the EDL shown strong influence of the surface potential and carrier density with positive gate voltage. Therefore, we performed gate-dependent PL spectra to identify exciton assignment. The Fig 3b shows PL spectra for different voltage configuration at room temperature. It is evident that no trion peak emerge in the whole range, and the PL intensity is higher at negative bias and shows a negative correlation with positive voltage. However, the position of exciton peak is not relative to gate voltage. Besides, in Fig 3c, we studied the excitation power-dependence PL. The peak intensities rapidly increase as excitation power increasing, while the peak position is still constant. We noted that new peak positions emerge when excitation power is 0.71 $\mu J \cdot cm^{-2}$. The emerged peak is lower than extion, which might be recognised as trion peak but the intensity is so week to be negligible[14]. Hence, we infer that the substrate can facilitate the formation of exciton. Those indicate that the SLGC and EDL at interface show extinct property on photonic dynamics comparing with traditional $SiO_2$. In order to identify the effect of EDL at interface, we introduced multilayer *h*-BN. After insertion of *h*-BN, Fig 3d displays the temperature-dependent PL spectra of h-BN inserted $WS_2$ and it experienced a similar blue shift with temperature decreasing. It is evident that the inserted h-BN have no weight on peak position, both peak position is same and valid. But *h*-BN facilitate extra electron hole pair generation and gain evident enhancement in intensity. Note that the full width at half maximum (FWHM) dramatically reduced in Fig 3e. Besides there is no trion peak even with high pump influence as displayed in Fig 3f. Since we do not observe a significant trion contribution in the PL spectra, we speculate that there might be strong interaction between carriers and defect[15]. We conclude that the substrate has positive effect on exciton formation and suppress the trion formation.

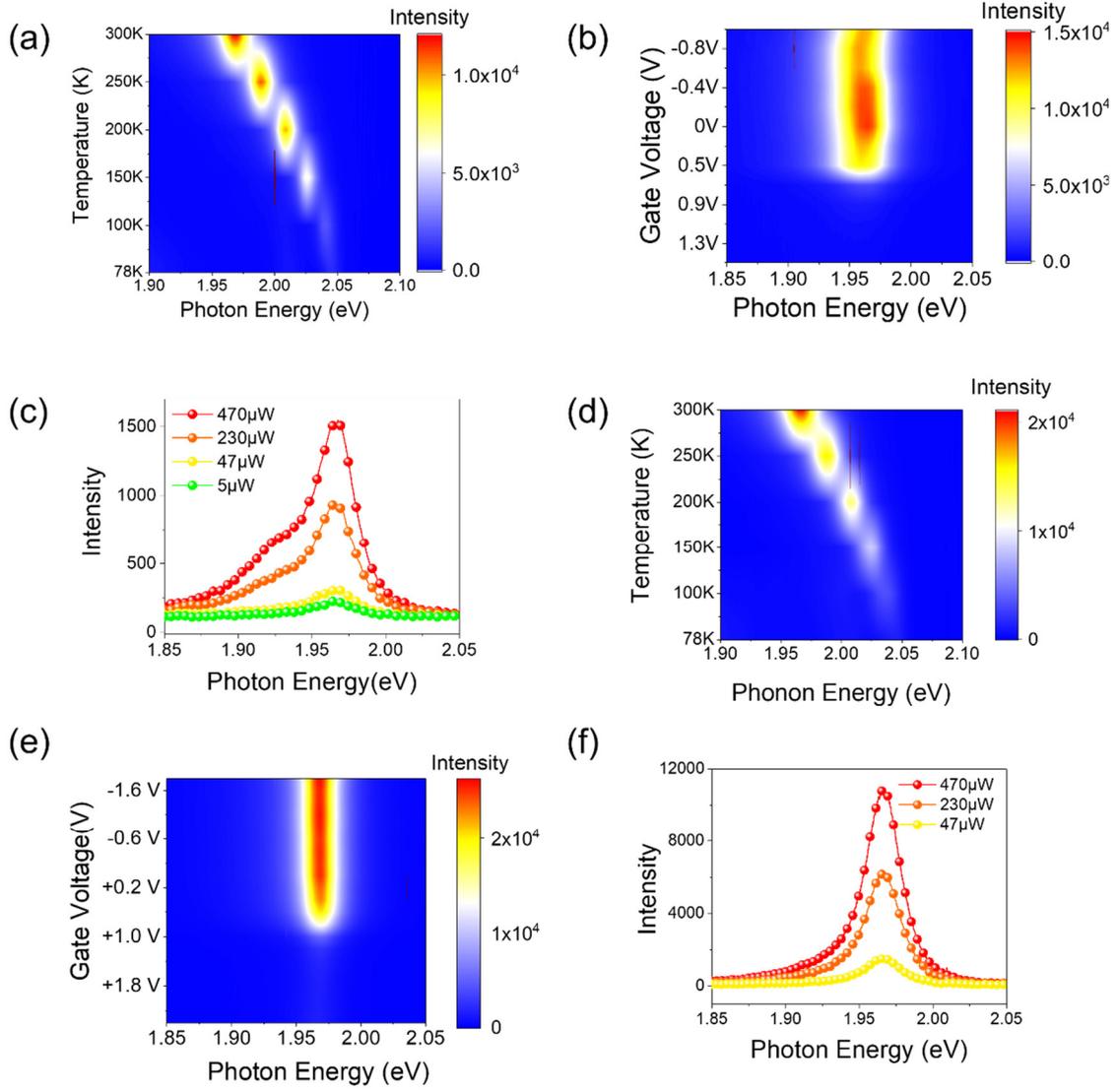

Figure 3. (a) The color plot of the temperature-dependent PL spectra for monolayer WS$_2$ and (d) h-BN inserted WS$_2$. (b) The gate dependent measurements for monolayer WS$_2$ and (e) h-BN inserted WS$_2$. (c) The excitation power dependent measurements for monolayer WS$_2$ and (f) h-BN inserted WS$_2$.

Furthermore, we pay attention to photonic dynamics. We established the PL decay of monolayer WS$_2$ on solid Lithium-ion glass substrate (SLGC) with systematic measurement of time-resolved PL spectra at voltage range of (0V ~ 1.2V), the result are shown in Fig 4. Since we kept the incident laser fluence at a relative low level ($0.06 \mu J \cdot cm^{-2}$), the nonequilibrium many-body effects such as Auger recombination can be neglected. For the data presented in Fig 4a, the PL decays can be fitted satisfactory with a double exponential decay convoluted with a Gaussian Instrument response function that is described by equation.

$$I(t) = \int_{-\infty}^{t} IRF \sum_{i=1}^{2} A_i \, e^{-\frac{t}{\tau_i}} dt$$

At different volage bias, we have best fitted the PL intensity with double-exponential decay behavior with two characteristic lifetimes including $\tau_1$, $\tau_2$, which implies that PL originates from a multiple state (other than only electron-hole recombination), indicate that other energy levels such as trap state induced by defects are significant in WS$_2$ on SLGC. It suggests that the introduction of solid electrolyte substrate might introduce some disorders between the surface of WS$_2$ and substrate[16-18], in case that WS$_2$ have a higher quantum yield compared with other TMDs and exhibit mono-exponential decay behavior on traditional SiO$_2$. Furthermore, with increasing voltage, the fast decay component becomes more significant, as attested by the prominent peak of TRPL traces at short delay times, whereas the slower decay component tends to decrease, with a corresponding smaller slope observed at larger delay times at high voltage. As the bias voltage increase from 0V to +1.2V, we fitted all the data measured with the double-exponential model, giving values of $\tau_1$ of thousands to hundreds nanoseconds and $\tau_2$ of hundreds to dozens nanoseconds as presented in Fig 4c. The average PL decay time can be calculated by the equation:

$$\tau_{ave} = \frac{A_1(\tau_1)^2 + A_2(\tau_2)^2}{A_1\tau_1 + A_2\tau_2}$$

According to equation, in Fig 4d, we obtain an average PL lifetime of 873 ps at zero bias, and the lifetime decrease with positive voltage, and result in 98 ps at +1.2 V. We attributed it to increasing carrier density induced by high gate voltage, there is more electrons, holes, and excitons, leading to a stronger interaction and heavier scattering among carriers which increase the probability for the relaxation of exciton. Therefore the decay process becomes faster. There have been a lot researches about the lifetime scale of each channel. We attribute the fast decay to trapping or defect-assisted recombination mechanisms, while the slow decay reflects the exciton radiative recombination[19, 20]. After insertion of multilayer h-BN, in Fig 4b, the exciton lifetime of exciton significantly extended at zero bias. According to equation, it is calculated that the exciton lifetime extend to ~2.89 ns which is about four times larger than original lifetime in the same configuration from the data of Fig 4d. And we suppose that the increase of the lifetime is due to the suppression of nonradiative recombination of excitons on the surface and interface defects, which is attributed to the dangling-bond-free h-BN surface[21]. They could suppress charge transfer and screen the doping effect, leading to the lower background carrier density or change of interlayer phonon modes. Thus, we conclude that the slower exciton lifetime is due to the screen effect of the h-BN layer. In Fig 4d, it is interesting that the lifetime quickly decreases to hundreds nanoseconds which is quantitively comparable with the pristine WS$_2$ as the gate voltage increased to +1.2V. Such phenomenon means that the screen effect of *h*-BN weakens on account of higher electric field. This further validate that the surface does exist electrical disorders which are sensitive to electric field and carry a big weight on exciton dynamics of WS$_2$. The observed disorders may depend on ions dynamics of EDL, which can be affected by high gate voltage.

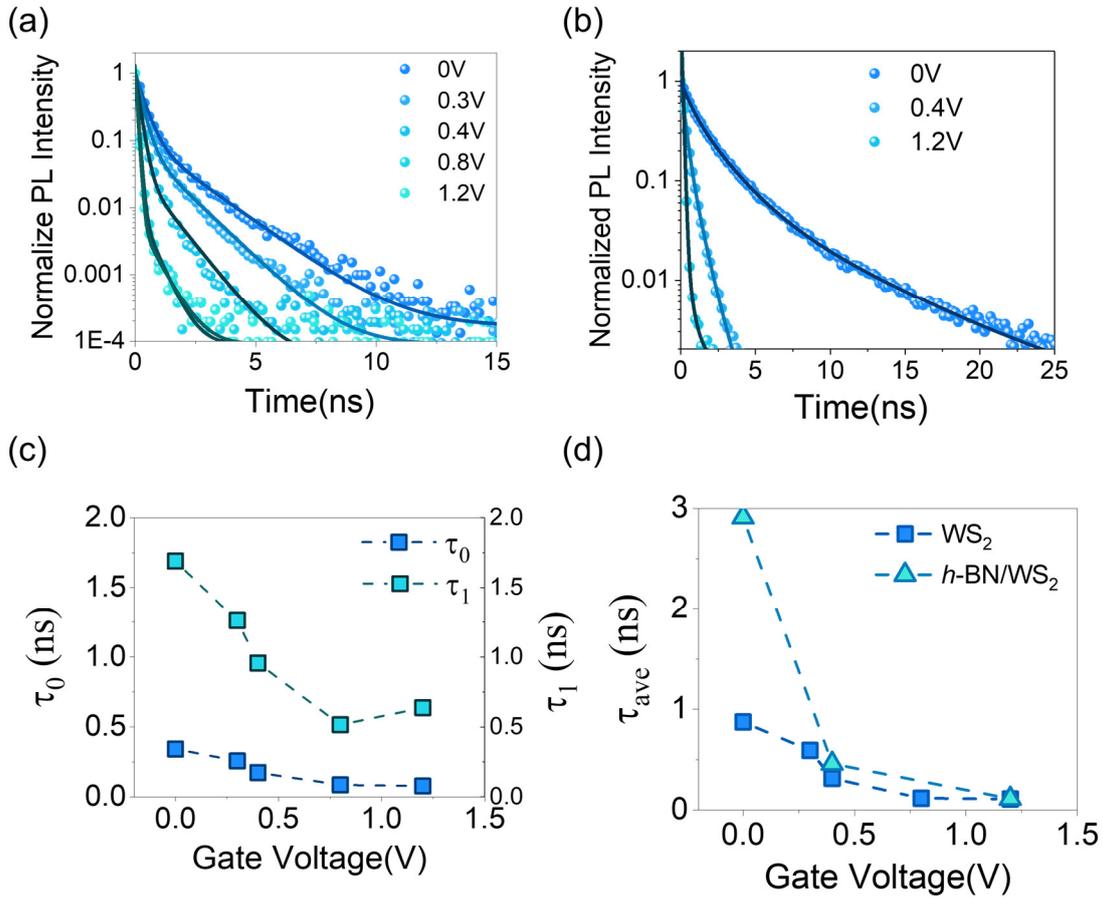

Figure 4. (a)Voltage-dependent TRPL plots for monolayer WS$_2$ and (b) *h*-BN inserted WS$_2$. (c) The voltage-dependent quick lifetime($\tau_0$) and the slow lifetime($\tau_1$) for monolayer WS$_2$ and (d) *h*-BN inserted WS$_2$.

We now focus on the exciton dynamics in the nonlinear regime. Exciton dynamics usually involve complex procedures with different recombination channels such as exciton-exciton annihilation, phonon scattering, electron-hole recombination and surface trapping. We took systematic pump fluence dependent time-resolved PL spectra to explore exciton dynamics at different exciton density, in terms of only radiation recombination or combined with non-radiation recombination. As shown in Fig 5a, the PL dynamics of the monolayer $WS_2$ is strongly pump density dependent. As pump density increase, additional shortening of the fast decay time is observed as evidenced by the faster initial decay. The fast dynamics can be attributed to Auger recombination (non-radiation recombination) or exciton-exciton annihilation, in case that two excitons interact with each other and fuse to form a higher energy then degenerate to the lowest energy excited state through energy conversion in the system[22]. It is noted that other mechanisms such as bi-exciton formation could occur at high pump fluence and contribute to fast dynamics[23]. However, bi-exciton formation would be accompanied by an additional low energy emission peak in PL spectra. For the range of pump fluence we used in this experiments, there is no bi-exciton peak even at the highest pump fluence. Therefore, we exclude the contribution of biexciton formation to the relevance of photonic dynamics observed here. As discussed before, exciton-exciton annihilation of $WS_2$ is a bimolecular process involving two excitons on account of strongly bound exciton and the contribution related to the formation of biexciton complexed has been ruled out. Instead, the proposed recombination mechanism which is responsible for the observed non-linear exciton dynamics is recognized as the exciton-exciton annihilation. In this framework, the annihilation rate equation can be written as:

$$\frac{dn}{dt} = -\frac{n}{\tau} - \gamma n^2$$

n, τ and γ represent the exciton population, the average exciton lifetime in the absence of exciton annihilation and the annihilation rate constant, respectively. Given that the γ is time-independent, the above equation is equal to following form:

$$\frac{1}{n(t)} = \left(\frac{1}{n(0)} - \gamma\tau\right) exp\left(\frac{t}{\tau}\right) - \gamma\tau$$

The monolayer $WS_2$ pump fluence dependent data are replotted in Fig 5b and Fig 5d. The equation is applied to make a global fit to the whole data to determine the exciton-exciton annihilation rate γ.

At zero bias configuration, for bare WS2, we obtain an exciton-exciton annihilation rate γ of 1.99 ± 0.27

× 10⁻³ cm² / s as the pump density of to 1.77 $\mu J \cdot cm^{-2}$. After insertion of h-BN, the annihilation rate of 5.15 ± 0.3 × 10⁻⁴ cm² / s is one order of magnitude lower comparing with bare WS2. Nevertheless, under highest pump fluence at positive voltage of +1.2V, both the annihilation rate reach the same value of 1.9 × 10⁻³ cm² / s, which implies that voltage is definitely to be the decisive impactor when the voltage is high enough. There are some questions needs to be discussed.

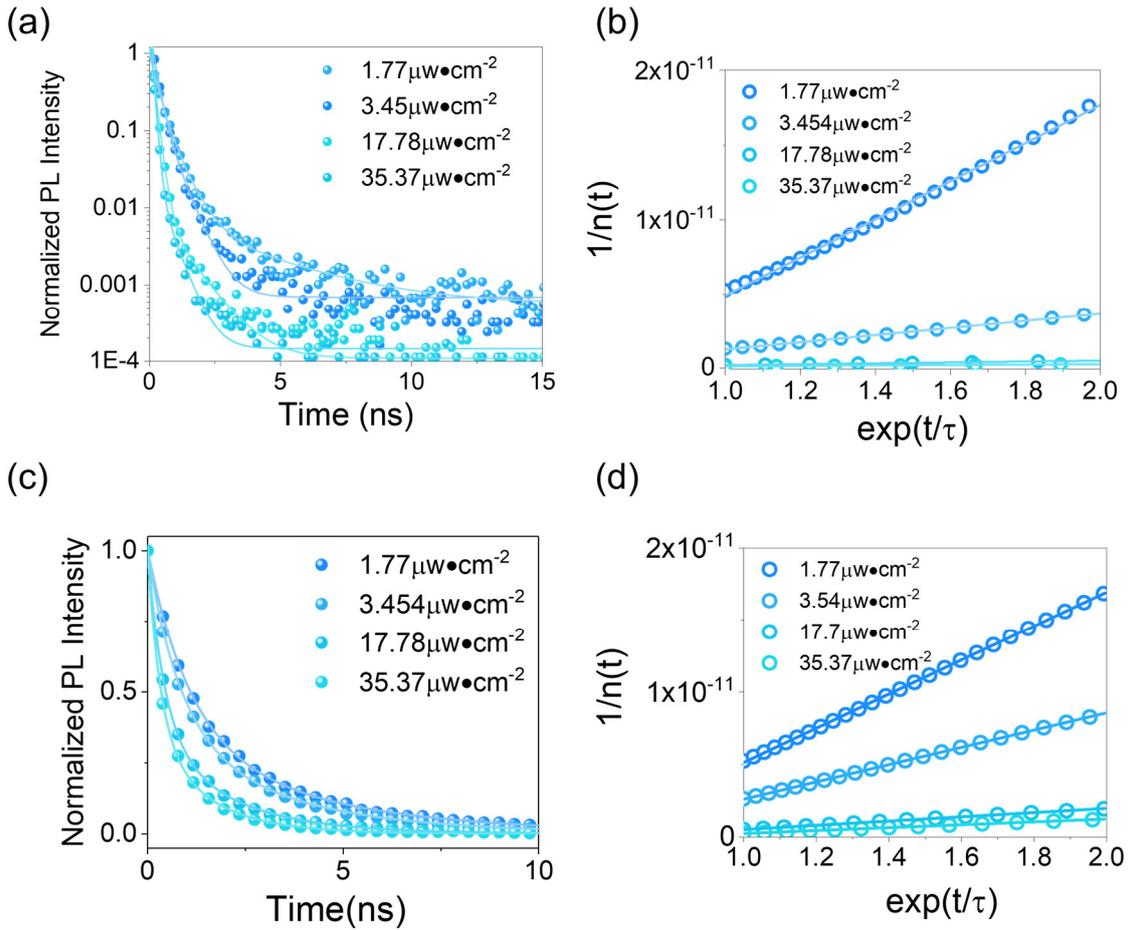

Figure 4. (a) PL decay at different excitation densities for monolayer WS$_2$ and (c) *h*-BN inserted WS$_2$. Linearized data for (b)monolayer WS$_2$ and (d) *h*-BN inserted WS$_2$

Firstly, the annihilation rate is relative lower than common value of 0.41±0.02 cm²/s in other papers. We explain the decrease of annihilation rate as follows. According to the Smuluchowski's work, the annihilation rate is related to the exciton diffusion constant and the maximum distance at which the annihilation process occurs through[24, 25]. That indicates that a lower exciton diffusivity leads to less exciton-exciton encounter, resulting in a lower overall annihilation rate. Meanwhile as depicted in Fig 3d, the full width at half maximum intensity is smaller after insertion of h-BN, which agree with weaken diffusion rate and further lead to decrease of annihilation. By the same taken, increasing of the voltage and pump fluence could give rise of FWHM equivalently and thus lead to increase of annihilation rate theoretically.

Secondly, it is evident that electrical field show strong influence on exciton lifetime, which shows positive trend with increasing gate voltage. We ascribe the discrepancy between two kinds sample to substrate-induced disorder. The defective surface system facilitate annihilation, and the inhomogeneous energy gradients exist as Auge annihilation centers, in which the excitons being trapped and are more likely to meet. In that case the defect-assisted Auger recombination rate will increase with increase of trap depth and trap density. At zero bias, the disorder of solid electrolyte substrate is critical factor. The insertion of BN significantly screens the disorder thus leading to lower exciton lifetime. However, the electric field induced by EDL at the interface gets stronger with higher voltage. In that case, there will be more carriers and the large amount of carriers become the dominant factor. Above all, we conclude that surface disorders play a dominant role at zero bias, diffusivity show more influence as voltage bias increasing in which situation that carrier density become pivotal factor.

**Conclusion**

We have identified that solid electrolyte is an ideal platform to gain insight into the EDL at interface, and avoid the disadvantage of liquid electrolyte. The solid electrolyte substrate shows strong gating ability in aspect of charged carrier immigration and distribution. We have demonstrated that the ion distribution of EDL is homogeneous through the interface according to uniform surface potential mapping. All the device shows quick response with voltage and have good stability after several test cycles. The time-resolved PL demonstrated that there is strong electric disorder induced by EDL, leading to slow immigration of exciton and suppress the annihilation. Our results have demonstrated the electricity dynamics, photonic dynamics of TMDCs EDL and proved that the solid electrolyte to be useful on exploring the interface reaction of EDL.